\documentclass{elsart}

\usepackage{epsfig}

\begin{document}

\def\d{{\rm d}}

\def\lp{\left. }
\def\rp{\right. }
\def\lr{\left( }
\def\rr{\right) }
\def\le{\left[ }
\def\re{\right] }
\def\lg{\left\{ }
\def\rg{\right\} }
\def\lb{\left| }
\def\rb{\right| }

\def\beq{\begin{equation}}
\def\eeq{\end{equation}}
\def\bea{\begin{eqnarray}}
\def\eea{\end{eqnarray}}

\begin{frontmatter}



\title{Slepton\,production\,in\,polarized\,hadron\,collisions}


\author{G.\ Bozzi, B.\ Fuks, and M.\ Klasen}

\address{Laboratoire de Physique Subatomique et de Cosmologie,
 Universit\'e Joseph Fourier/CNRS-IN2P3, 53 Avenue des Martyrs,
 F-38026 Grenoble, France}

\begin{abstract}
We calculate cross sections and asymmetries for slepton pair production
through neutral and charged electroweak currents in polarized hadron
collisions for general slepton masses and including mixing of the left- and
right-handed interaction eigenstates relevant for third generation sleptons.
Our analytical results confirm and extend a previous calculation.
Numerically, we show that measurements of the longitudinal single-spin
asymmetry at the existing polarized $pp$ collider RHIC and at possible
polarization upgrades of the Tevatron or the LHC would allow for a
determination of the tau slepton mixing angle and/or the associated
supersymmetry breaking parameters $\Lambda$ for gauge mediation and $A_0$
for minimal supergravity. Furthermore, the Standard Model background from
tau pair production can be clearly distinguished due to the opposite sign of
the associated asymmetry.
\end{abstract}

\end{frontmatter}


\vspace*{-155mm}
LPSC 04-091
\vspace*{ 140mm}

\section{Introduction}
\label{sec:1}

One of the most promising extensions of the Standard Model (SM) of particle
physics is the Minimal Supersymmetric Standard Model (MSSM)
\cite{Nilles:1983ge,Haber:1984rc}, which postulates a symmetry between
fermionic and bosonic degrees of freedom in nature and predicts the
existence of a fermionic (bosonic) supersymmetric (SUSY) partner for each
bosonic (fermionic) SM particle. Since SUSY and SM particles contribute to
the quadratic divergence of the mass of the Higgs boson with equal strength,
but opposite sign, the MSSM can, {\it inter alia}, stabilize the electroweak
energy scale with respect to the Planck scale and thus propose a solution to
the hierarchy problem.

Unfortunately, SUSY particles still remain to be discovered. Their masses
must therefore be considerably larger than those of the corresponding SM
particles, and the symmetry is bound to be broken. In order to remain a
viable solution to the hiearchy problem, SUSY can, however, only be
broken via soft mass terms in the Lagrangian, with the consequence that the
SUSY particle masses should lie in the TeV range and thus within the
discovery reach of current and future hadron colliders such as the Tevatron
and the LHC.

Production cross sections for SUSY particles at hadron colliders have been
extensively studied in the past at leading order (LO) \cite{Dawson:1983fw,%
delAguila:1990yw,Baer:1993ew} and also at next-to-leading order (NLO) of
perturbative QCD \cite{Beenakker:1996ch,Beenakker:1997ut,Berger:1998kh,%
Berger:1999mc,Berger:2000iu,Baer:1997nh,Beenakker:1999xh}. In particular,
the QCD \cite{Baer:1997nh} and full SUSY-QCD \cite{Beenakker:1999xh}
corrections for slepton pair production are known to increase the hadronic
cross sections by about 35 \% at the Tevatron and 25\% at the LHC, thus
extending their discovery reaches by several tens of GeV.

Despite the first successful runs of the RHIC collider in the polarized $pp$
mode, polarized SUSY production cross sections have received much less
attention. Only the pioneering LO calculations for massless squark and
gluino production \cite{Craigie:1983as,Craigie:1984tk} have recently been
confirmed, extended to the massive case, and applied to current hadron
colliders \cite{Gehrmann:2004xu}.

It is the aim of this work to verify the corresponding pioneering LO
calculation for slepton pair production \cite{Chiappetta:1985ku} and include
the mixing effects relevant for third generation sleptons. Our analytical
results for neutral ($\gamma$, $Z^0$) and charged ($W^\pm$) current
slepton and lepton pair production will be presented in Sec.\ \ref{sec:2}.
In Sec.\ \ref{sec:3}, numerical predictions will be made for unpolarized
cross sections and longitudinal spin asymmetries at RHIC and possible
upgrades of the Tevatron \cite{Baiod:1995eu} and the LHC \cite{roeck}.
Particular emphasis will be put on the
sensitivity of the asymmetry to the tau slepton mixing angle as predicted by
various modern SUSY breaking mechanisms \cite{Allanach:2002nj,spa}.
Possibilities to discriminate between the SUSY signal and the Drell-Yan SM
background will also be discussed. We summarize our results in Sec.\
\ref{sec:4}.

\section{Analytical results}
\label{sec:2}

In order to be able to compare directly with the previously published LO
cross sections for slepton pair production in unpolarized hadron collisions
\cite{Dawson:1983fw}, we define the square of the weak coupling constant
$g_W^2=e^2/\sin^2\theta_W$ in terms of the electromagnetic fine structure
constant $\alpha=e^2/(4\pi)$ and the squared sine of the electroweak mixing
angle $x_W=\sin^2\theta_W$. The coupling strengths of left- and right-handed
(s)fermions to the neutral electroweak current are then given by
\bea
 L_f = 2\,T^{3}_f-2\,e_f\,x_W &\mbox{~~and~~}&
 R_f =           -2\,e_f\,x_W,
\eea
where the weak isospin quantum numbers are $T_f^3=\pm1/2$ for left-handed
and $T_f^3=0$ for right-handed up- and down-type (s)fermions, and their
fractional electromagnetic charges are denoted by $e_f$.

In general SUSY breaking models, where the sfermion interaction eigenstates
are not identical to the respective mass eigenstates, the coupling strenghts
$L_f$ and $R_f$ must be multiplied by $S_{j1}S_{i1}^\ast$ and $S_{j2}
S_{i2}^\ast$, respectively, where $i,j\in\{1,2\}$ label the sfermion mass
eigenstates (conventionally $m_{\tilde{f}_1} < m_{\tilde{f}_2}$) and $S$
represents the unitary matrix diagonalizing the sfermion mass matrix
(see App.\ \ref{sec:a}). Including these slepton mixing effects in the
polarized cross sections for the production of slepton pairs in hadron
collisions represents our main analytical improvement over the previously
published results in Ref.\ \cite{Chiappetta:1985ku}.

Our results for the electroweak $2\to2$ scattering process
\bea
 q_{h_a}(p_a)\overline{q}_{h_b}(p_b)\to\tilde{l}_i(p_1)\tilde{l}^*_j(p_2)
\eea
will be expressed in terms of the conventional Mandelstam variables,
\bea
 s=(p_a+p_b)^2 &~~,~~& t=(p_a-p_1)^2 \mbox{~~,~~and~~} u=(p_a-p_2)^2
\eea
and the masses of the neutral and charged electroweak gauge bosons $m_Z$
and $m_W$.

\subsection{Sleptons}

The neutral current cross section for the production of non-mixing slepton
pairs in collisions of quarks with definite helicities $h_{a,b}$ is given
by
\bea
 {\d\hat{\sigma}_{h_a,h_b}\over\d t} &=& {4 \pi \alpha^2 \over 3 s^2}
 \le {u t - m_i^2 m_j^2 \over s^2} \re 
     \le e_q^2 e_l^2 (1 - h_a h_b) \rp \nonumber \\
 &+& {e_q e_l (L_l + R_l) [(1-h_a) (1+h_b) L_q + (1+h_a) (1-h_b) R_q]
     \over 8 x_W (1 - x_W) (1-m_Z^2/s)} \nonumber \\
 &+& \lp {(L_l^2 + R_l^2) [(1-h_a) (1+h_b) L_q^2 + (1+h_a) (1-h_b) R_q^2]
     \over 64 x_W^2 (1 - x_W)^2 (1-m_Z^2/s)^2} \re. \label{eq:4}
\eea
For the production of the slepton mass eigenstates $i$ and $j$, the
couplings $L_l$ and $R_l$ must be modified as described above. In addition,
the first two lines in Eq.\ (\ref{eq:4}), representing the squared photon
and photon-$Z^0$ interference contributions, receive additional factors of
$\delta_{ij}/2$ and $\delta_{ij}$, respectively, while in the third line,
representing the squared $Z^0$ contribution, the factor $L_l^2+R_l^2$ must
be replaced by $(L_l+R_l)^2$.

The pure left-handed, charged current cross section
\bea
 {\d\hat{\sigma}_{h_a,h_b}\over\d t} &=& {4 \pi \alpha^2 \over 3 s^2}
 \le {u t - m_i^2 m_j^2 \over s^2} \re
 \le {(1 - h_a) (1 + h_b) \over 16 x_W^2 (1 - m_W^2/s)^2} \re
\eea
is easily derived from Eq.\ (\ref{eq:4}) by setting
\bea
 m_Z~\rightarrow~m_W,~~~
 e_q=e_l=R_q=R_l=0, &\mbox{~~~and~~~}&
 L_q=L_l=\sqrt{2}\cos\theta_W.
 \label{eq:6}
\eea

Averaging over initial helicities,
\bea
 \d\hat{\sigma}&=&
 {\d\hat{\sigma}_{ 1, 1}
 +\d\hat{\sigma}_{ 1,-1}
 +\d\hat{\sigma}_{-1, 1}
 +\d\hat{\sigma}_{-1,-1}
 \over 4},
\eea
we obtain the unpolarized partonic cross section
\bea
 {\d\hat{\sigma}         \over\d t} &=& {4 \pi \alpha^2 \over 3 s^2}
 \le {u t - m_i^2 m_j^2 \over s^2} \re \\
 &\times& \le e_q^2e_l^2
 + {e_q e_l (L_q + R_q) (L_l + R_l) \over  8 x_W   (1-x_W)   (1-m_Z^2/s)}
 + {(L_q^2 + R_q^2) (L_l^2 + R_l^2) \over 64 x_W^2 (1-x_W)^2 (1-m_Z^2/s)^2}
 \re, \nonumber
\eea
which agrees for non-mixing sleptons with the neutral current result of
Ref.\ \cite{Dawson:1983fw} in the limit of equal masses $m_L=m_R$ and with
the charged current result of Ref.\ \cite{Baer:1993ew}. Note that for
invariant final state masses close to the $Z^0$-pole, $s\simeq m_Z^2$, the
$Z^0$-propagators must be modified to include the decay width of the $Z^0$
boson.

From Eq.\ (\ref{eq:4}), one can easily calculate the longitudinal
double-spin asymmetry $A_{LL}$ using the polarized differential cross
section
\bea
 \d\Delta\hat{\sigma}_{LL}&=&
 {\d\hat{\sigma}_{ 1, 1}
 -\d\hat{\sigma}_{ 1,-1}
 -\d\hat{\sigma}_{-1, 1}
 +\d\hat{\sigma}_{-1,-1}
 \over 4}.
\eea
However, the result
\bea
 A_{LL} &=& {\d\Delta\hat{\sigma}_{LL}\over\d\hat{\sigma}} = -1
\eea
is totally independent of all SUSY breaking parameters.

It will thus be far more interesting to calculate the single-spin asymmetry
$A_L=\d\Delta\hat{\sigma}_L/\d\hat{\sigma}$ from the polarized differential
cross section
\bea
 \d\Delta\hat{\sigma}_L&=&
 {\d\hat{\sigma}_{ 1, 1}
 +\d\hat{\sigma}_{ 1,-1}
 -\d\hat{\sigma}_{-1, 1}
 -\d\hat{\sigma}_{-1,-1}
 \over 4},
\eea
{\it i.e.} for the case of only one polarized hadron beam. Not only does the
neutral current cross section
\bea
 \d\Delta\hat{\sigma}_L&=& {4 \pi \alpha^2 \over 3 s^2}
 \le u t - m_i^2 m_j^2 \over s^2 \re \label{eq:12} \\
 &\times&
 \le - {e_q e_l (L_l + R_l) (L_q - R_q) \over
       8 x_W (1 - x_W) (1 - m_Z^2/s)}
     - {(L_l^2 + R_l^2) (L_q - R_q) (L_q + R_q) \over
       64 x_W^2 (1 - x_W)^2 (1 - m_Z^2/s)^2} \re \nonumber
\eea
remain sensitive to the SUSY breaking parameters, but even more the
squared photon contribution, which is insensitive to these parameters, is
eliminated. Finally, this scenario may also be easier to implement
experimentally, {\it e.g.} at the Tevatron, since protons are much more
easily polarized than antiprotons \cite{Baiod:1995eu}.

To conclude our analytical calculation of the polarized partonic slepton
cross sections, we note that our neutral current result in Eq.\
(\ref{eq:12}) as well as our charged current result
\bea
 \d\Delta\hat{\sigma}_L&=& {4 \pi \alpha^2 \over 3 s^2}
 \le u t - m_i^2 m_j^2 \over s^2 \re
 \le {-1 \over 16 x_W^2 (1 - m_W^2/s)^2} \re
\eea
agree \cite{footnote} with those in Ref.\ \cite{Chiappetta:1985ku} for
non-mixing sleptons after integration over $t$ in the interval
\bea
 t_{\min,\max} &=&
 -{s+m_j^2-m_i^2\over2}\mp{\sqrt{(s-m_i^2-m_j^2)^2-4 m_i^2 m_j^2}\over2}
 +m_j^2. \label{eq:14}
\eea

\subsection{Leptons}

Due to their purely electroweak couplings, sleptons are among the lightest
SUSY particles in many SUSY breaking scenarios \cite{Allanach:2002nj,spa}
and often decay directly into the stable lightest SUSY particle (LSP), which
may be the lightest neutralino in minimal supergravity (mSUGRA) models or
the gravitino in gauge mediated SUSY breaking (GMSB) models. The slepton
signal at hadron colliders will therefore consist in a lepton pair, which
will be easily detectable, and associated missing (transverse) energy. This
forces us to consider also the corresponding background of SM lepton pair
production through Drell-Yan type processes.

With the help of the mass-subtracted Mandelstam variables
\bea
 t_{i,j}~=~t-m_{i,j}^2 &\mbox{~~and~~}& u_{i,j}~=~u-m_{i,j}^2,
\eea
we can write the polarized neutral current cross section as
\bea
 {\d\hat{\sigma}_{h_a,h_b}\over\d t} &=& {4 \pi \alpha^2 \over 3 s^2}
 \le e_q^2 e_l^2 (1 - h_a h_b)
 {s (m_i + m_j)^2 + t^2 + u^2-m_i^4 - m_j^4 \over 2 s^2}\rp \\
 &+& e_q e_l
 {L_l [(1-h_a)(1+h_b) L_q (u_i u_j+m_i m_j s)]
 +R_l [t\leftrightarrow u]
 \over 8 x_W (1 - x_W) s (s-m_Z^2)} \nonumber \\
 &+& e_q e_l
 {L_l [(1+h_a)(1-h_b) R_q (t_i t_j+m_i m_j s)]
 +R_l [t\leftrightarrow u]
 \over 8 x_W (1 - x_W) s (s-m_Z^2)} \nonumber \\
 &+&
 {L_l^2 [(1-h_a)(1+h_b) L_q^2 u_i u_j+(1+h_a)(1-h_b) R_q^2 t_i t_j]
 +R_l^2 [t\leftrightarrow u]
 \over 64 x_W^2 (1 - x_W)^2 (s - m_Z^2)^2} \nonumber \\
 &+& \lp
 {2 L_l R_l m_i m_j s [(1-h_a)(1+h_b) L_q^2+(1+h_a)(1-h_b) R_q^2]
 \over 64 x_W^2 (1 - x_W)^2 (s - m_Z^2)^2} \re. \nonumber
\eea
From this equation, the charged current cross section
\bea
 {\d\hat{\sigma}_{h_a,h_b}\over\d t} &=& {4 \pi \alpha^2 \over 3 s^2}
 {(1-h_a) (1+h_b) u_i u_j \over  16 x_W^2 (s-m_W^2)^2}
\eea
can be obtained using again the substitions given in Eq.\ (\ref{eq:6}). For
polarized quarks and unpolarized, massless leptons our results agree with
Ref.\ \cite{Craigie:1984tk} and, up to an overall sign in the analytical
single-spin asymmetry, also with Ref.\ \cite{Chiappetta:1984dy}. After
averaging over initial spins and integration over $t$, our results also
agree with Ref.\ \cite{Ellis:1991qj}.

\section{Numerical Results}
\label{sec:3}

For the masses and widths of the electroweak gauge bosons, we use the
current values of $m_Z=91.1876$ GeV, $m_W=80.425$ GeV, $\Gamma_Z=2.4952$
GeV, and $\Gamma_W=2.124$ GeV. The squared sine of the electroweak mixing
angle
\bea
 \sin^2\theta_W &=& 1-m_W^2/m_Z^2
\eea
and the electromagnetic fine structure constant
\bea
 \alpha &=& \sqrt{2} G_F m_W^2 \sin^2\theta_W / \pi
\eea
can then be calculated in the improved Born approximation using the world
average value of $G_F=1.16637\cdot 10^{-5}$ GeV$^{-2}$ for Fermi's
coupling constant \cite{Eidelman:2004wy}.

Since the mixing of left- and right-handed slepton interaction eigenstates
is proportional to the mass of the corresponding SM partner (see App.\
\ref{sec:a}), it is numerically only important for third generation
sleptons. Consequently, the lightest slepton is the lighter stau mass
eigenstate $\tilde{\tau}_1$ in most SUSY breaking models
\cite{Allanach:2002nj,spa}, and we focus our numerical studies on its
production.

The mass limits imposed by the four LEP experiments on the tau slepton
vary between 52 and 95.9 GeV. They depend strongly on the assumed SUSY
breaking meachanism, the mass difference between the stau and the LSP, and
the stau mixing angle. The weakest limit of 52 GeV is found for GMSB models
and stau decays to gravitinos, if no constraints on their mass difference
are imposed \cite{Barate:1998zp}. This is the scenario that we will study
for the RHIC collider, which has the most restricted hadronic center-of-mass
energy range ($\sqrt{S}\leq 500$ GeV). For the Tevatron ($\sqrt{S}=1.96$
TeV) and at the LHC ($\sqrt{S}=14$ TeV) with their considerably larger
center-of-mass energies, we will, however, impose the stricter mass limit
of 81.9 GeV \cite{Abdallah:2003xe}, which is valid for stau decays to
neutralinos with a mass differences of at least 15 GeV and represents the
current standard value \cite{Eidelman:2004wy}. The SM background will be
evaluated using the physical tau mass of $m_\tau=1.77699$ GeV.

Our numerical calculations of cross sections and asymmetries for the current
(RHIC, Tevatron) and future (LHC) hadron colliders with up-to-date
parton densities represent the main numerical improvement of this work over
the previously published results in Ref.\ \cite{Chiappetta:1985ku}, which
discussed only the case of the CERN $Sp\overline{p}S$ collider at
$\sqrt{S}=540$ GeV with nowadays obsolete parton densities.

\subsection{Unpolarized cross sections for non-mixing sleptons}

Thanks to the QCD factorization theorem, unpolarized hadronic cross sections
\bea
 \sigma &~=&
 \int_{m^2/S}^1\!\d\tau\!\!
 \int_{-1/2\ln\tau}^{1/2\ln\tau}\!\!\d y
 \int_{t_{\min}}^{t_{\max}} \d t \
 f_{a/A}(x_a,M_a^2) \ f_{b/B}(x_b,M_b^2) \ {\d\hat{\sigma}\over\d t}
\eea
can be calculated by convoluting the relevant partonic cross section
$\hat{\sigma}$ with universal parton densities $f_{a/A}$ and $f_{b/B}$
of partons $a,b$
in the hadrons $A,B$, which depend on the longitudinal momentum fractions of
the two partons $x_{a,b} = \sqrt{\tau}e^{\pm y}$ and on the unphysical
factorization scales $M_{a,b}$. In order to employ a consistent set of
unpolarized and polarized parton densities (see below), we choose the LO set
of GRV98 \cite{Gluck:1998xa} for our unpolarized predictions at the
factorization scale $M_a=M_b=m=(m_i+m_j)/2$.

\begin{figure}
 \centering
 \epsfig{file=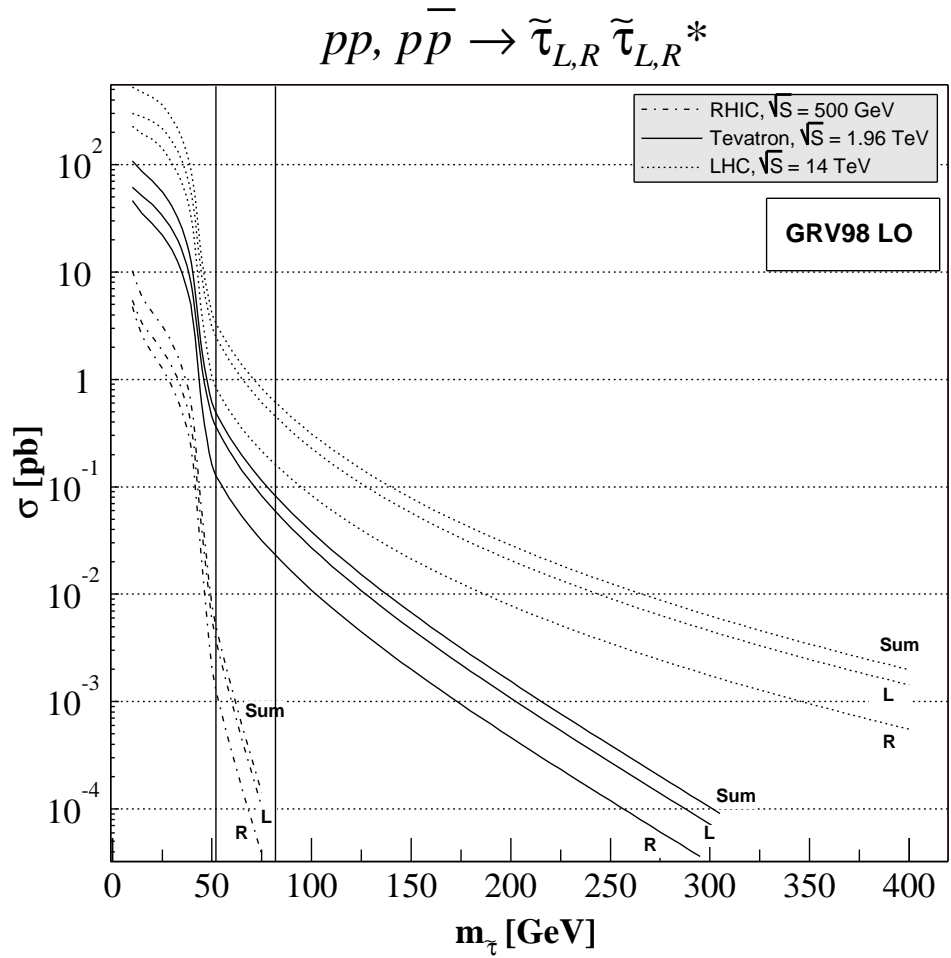,width=\columnwidth}
 \caption{\label{fig:1}Unpolarized hadronic cross sections for pair
 production of non-mixing tau sleptons at the RHIC, Tevatron, and LHC
 colliders as a function of their physical mass. For consistency with the
 polarized cross sections (see below), GRV98 LO parton densities have been
 used. The vertical lines indicate the two different stau mass limits of 52
 \cite{Barate:1998zp} and 81.9 GeV \cite{Abdallah:2003xe}.}
\end{figure}

In Fig.\ \ref{fig:1}, we show the unpolarized hadronic cross sections for
%
%
pair production of non-mixing tau sleptons at the RHIC, Tevatron, and LHC
colliders as a function of their physical mass. Unfortunately, the
observation of tau sleptons, as that of any SUSY particles, will be
difficult at RHIC, which is the only existing polarized hadron collider.
In contrast, tau sleptons will be detectable at the LHC over a large region
of the viable SUSY parameter space up to stau masses of about 400 GeV. At
the Tevatron, the discovery reach extends considerably beyond the current
exclusion limits.

We have checked that the unpolarized cross sections change by at most 10
\% if calculated with the more recent parton densities CTEQ6L1
\cite{Pumplin:2002vw}. Since the (sizeable) variations of the hadronic
cross sections with the unknown factorization scale at LO are considerably
reduced at NLO \cite{Baer:1997nh,Beenakker:1999xh} and cancel to a large
extent in the asymmetries, we refer the reader to these references for
detailed estimates of factorization scale uncertainties.

Before application of any experimental cuts, the SUSY signal cross sections
in Fig.\ \ref{fig:1} are at least three orders of magnitude smaller than the
corresponding SM background cross sections from tau lepton pair production
(1.7, 3.4, and 8.3 nb for the RHIC, Tevatron, and LHC colliders,
respectively, using GRV98 LO parton densities at $M_a=M_b=m_\tau$). Imposing
an invariant mass cut on the observed lepton pair and a minimal missing
transverse energy will, however, greatly improve the signal-to-background
ratio. In addition, as we will see in the next section, asymmetries may
provide an important tool to further distinguish the SUSY signal from the SM
background.

\subsection{Single-spin asymmetries for mixing sleptons}

Using again the QCD factorization theorem, we calculate the hadronic
cross section for longitudinally polarized hadrons $A$ with unpolarized
hadrons $B$
\bea
 \Delta\sigma_L &~=&
 \int_{m^2/S}^1\!\d\tau\!\!
 \int_{-1/2\ln\tau}^{1/2\ln\tau}\!\!\d y
 \int_{t_{\min}}^{t_{\max}} \d t \
 \Delta f_{a/A}(x_a,M_a^2) \ f_{b/B}(x_b,M_b^2) \ {\d\Delta\hat{\sigma}_L
 \over\d t}
\eea
through a convolution of polarized ($\Delta f_{a/A}$) and unpolarized
($f_{b/B}$) parton densities with the singly polarized partonic cross
section $\Delta\hat{\sigma}_L$.

As mentioned above, we employ a consistent set of unpolarized
\cite{Gluck:1998xa} and polarized \cite{Gluck:2000dy} LO parton densities.
We estimate the theoretical uncertainty due to the less well known polarized
parton densities by showing our numerical predictions for both the GRSV2000
LO standard (STD) and valence (VAL) parameterizations. Although these parton
densities differ from the older parton densities employed in Ref.\
\cite{Chiappetta:1985ku}, we have checked that our numerical predictions for
asymmetries of non-mixing sleptons and leptons are in reasonable agreement
with Fig.\ 1 of Ref.\ \cite{Chiappetta:1985ku}.

%
\begin{figure}
 \centering
 \epsfig{file=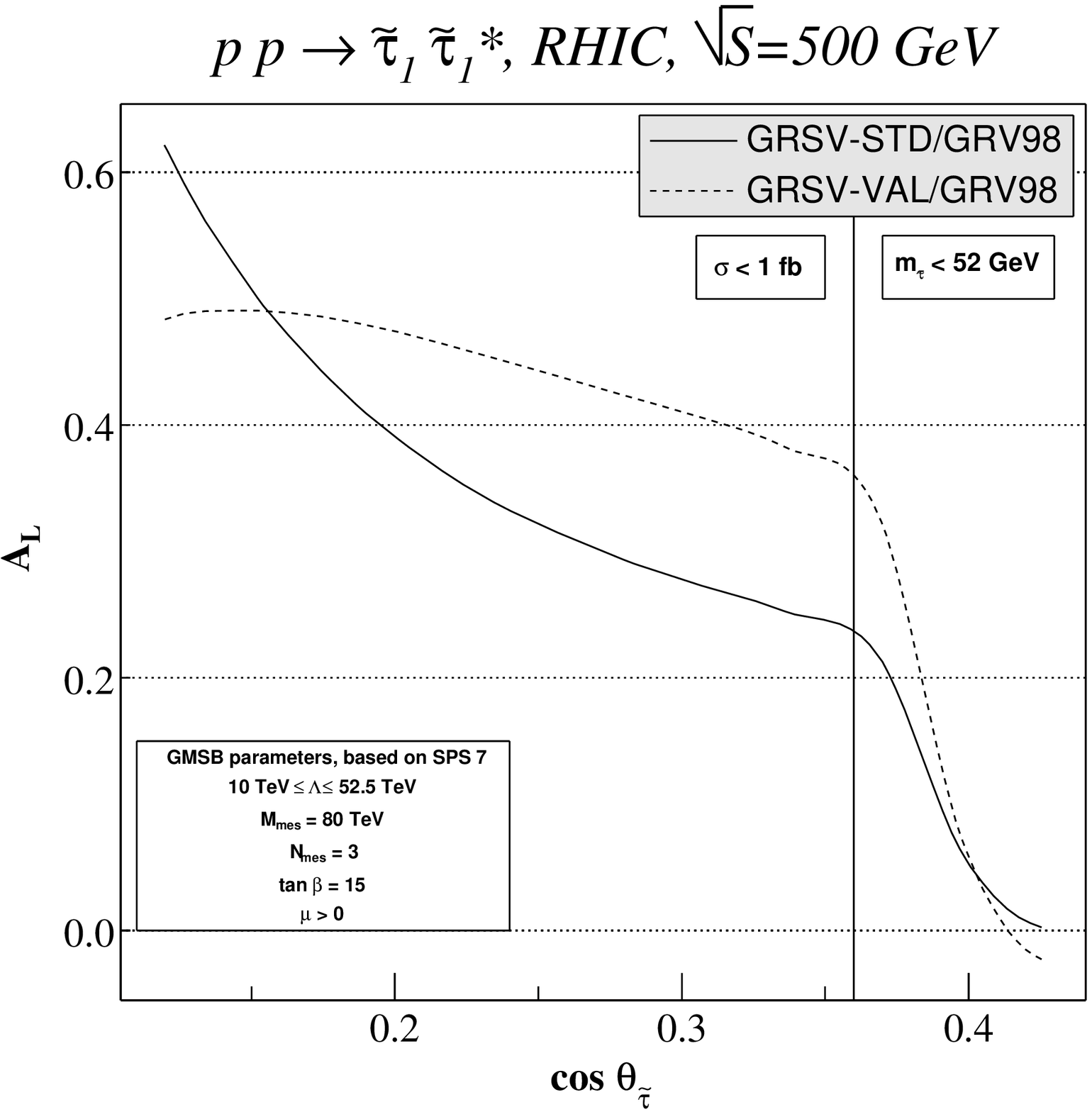,width=\columnwidth}
 \caption{\label{fig:2}Dependence of the longitudinal single-spin asymmetry
 $A_L$ on the cosine of the stau mixing angle for $\tilde{\tau}_1$ pair
 production in a GMSB model at RHIC. Although the asymmetry is large and
 depends strongly on the stau mixing angle, its determination will be
 difficult due to the limited center-of-mass energy and luminosity at RHIC.}
\end{figure}
%

Since we are primarily interested in the possible impact of slepton pair
production with polarized hadron beams on the determination of the mixing
angle for third generation sleptons in realistic SUSY breaking scenarios,
we choose for the three hadron colliders three of the ten benchmark points
introduced in Ref.\ \cite{Allanach:2002nj}: The GMSB point SPS 7 with a
light tau slepton decaying to a gravitino for RHIC and its very limited mass
range, the typical mSUGRA point SPS 1a' with an intermediate value of
$\tan\beta=10$ and a slightly reduced common scalar mass of $m_0=70$ GeV
\cite{spa} for the Tevatron, and the mSUGRA point SPS 4 with a large scalar
mass of $m_0=400$ GeV and large $\tan\beta=50$, which enhances mixing for
tau sleptons, for the LHC with its larger mass range.

%
\begin{figure}
 \centering
 \epsfig{file=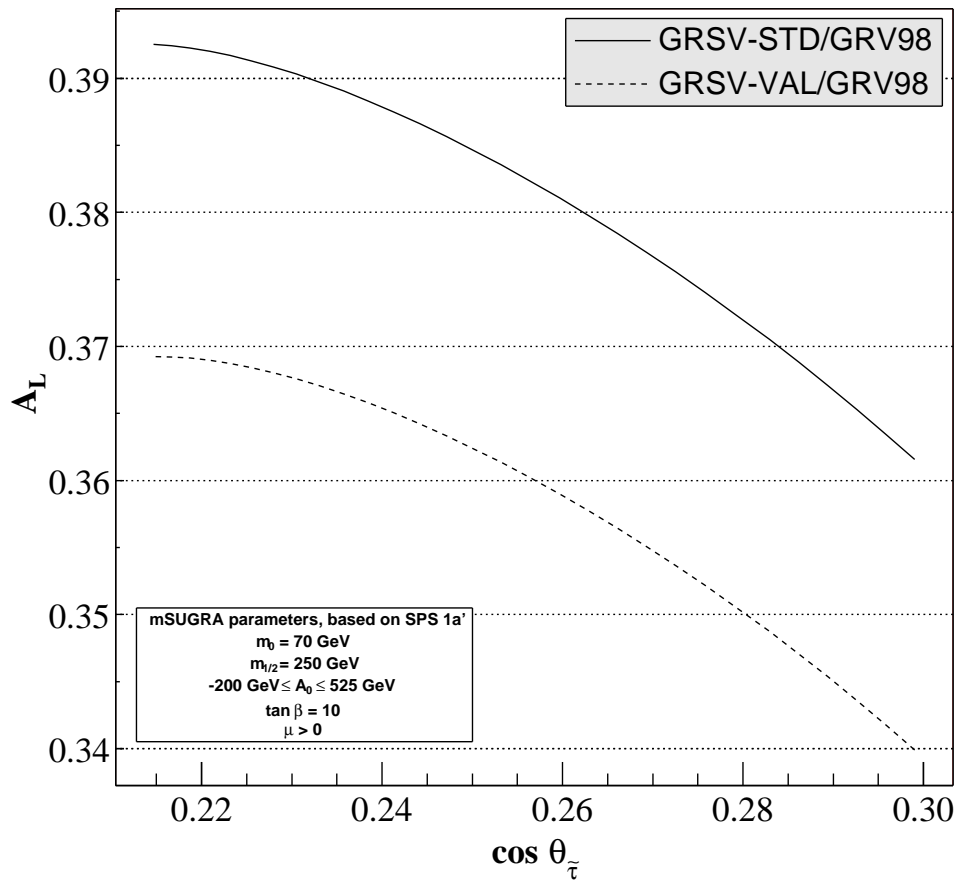,width=\columnwidth}
 \caption{\label{fig:3}Dependence of the longitudinal single-spin asymmetry
 $A_L$ on the cosine of the stau mixing angle for $\tilde{\tau}_1$ pair
 production in a typical mSUGRA model at the Tevatron. While the unpolarized
 cross section is nearly independent of $A_0$ and the stau mixing angle,
 these parameters could well be determined in an asymmetry measurement,
 provided the polarized parton densities are better constrained.}
\end{figure}
%

The physical mass of the pair produced light tau slepton mass eigenstate and
the mixing angle are calculated using the recently updated computer program
SUSPECT \cite{Djouadi:2002ze}. Its Version 2.3 includes now a consistent
calculation of the Higgs mass, with all one-loop and the dominant two-loop
radiative corrections, in the renormalization group equations, that link the
restricted set of SUSY breaking parameters at the gauge coupling unification
scale to the complete set of observable SUSY masses and mixing angles at the
electroweak scale.

\begin{figure}
 \centering
 \epsfig{file=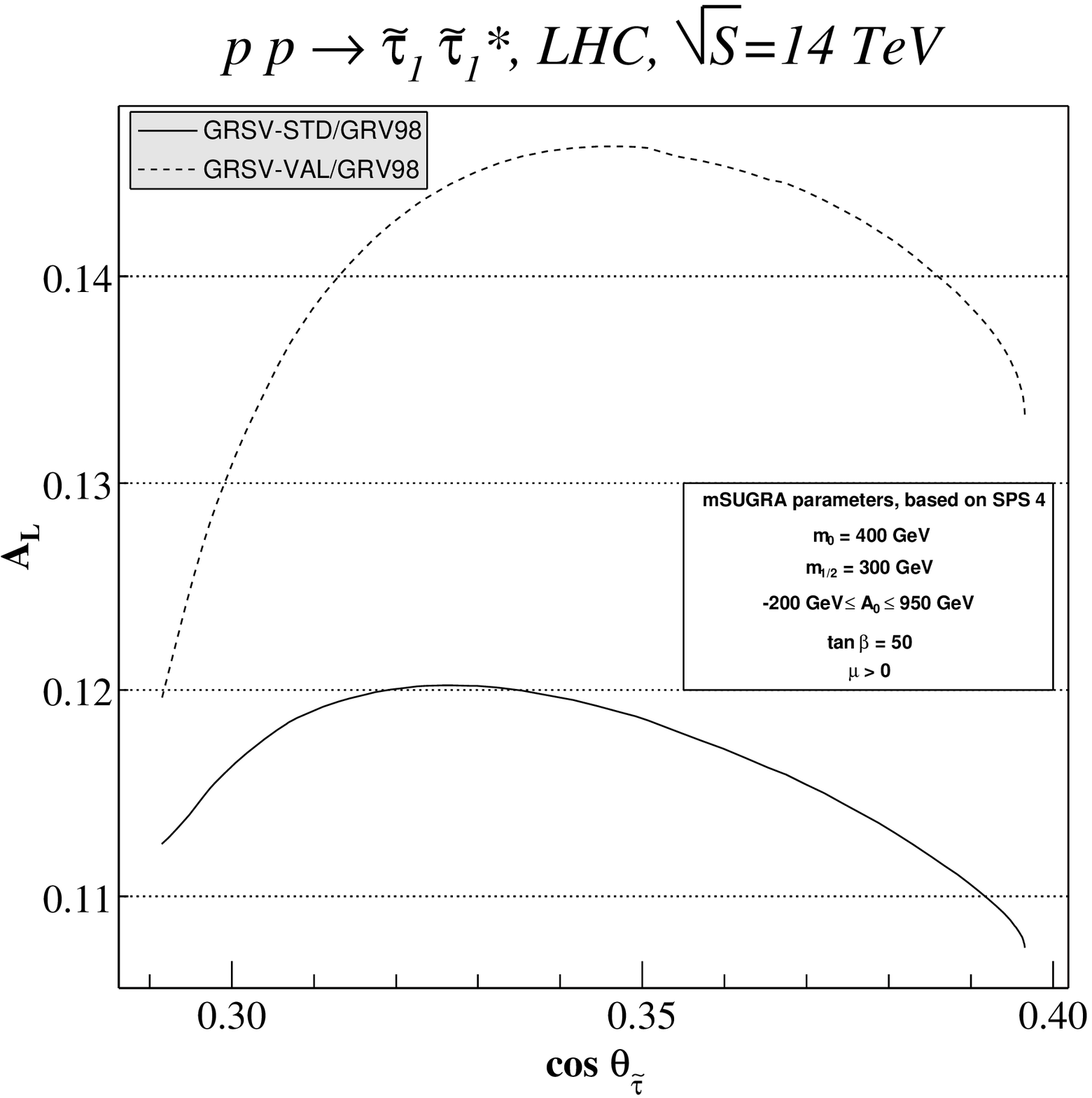,width=\columnwidth}
 \caption{\label{fig:4}Dependence of the longitudinal single-spin asymmetry
 $A_L$ on the cosine of the stau mixing angle for $\tilde{\tau}_1$ pair
 production in an mSUGRA model with large $m_0$ and $\tan\beta$ at the LHC.
 A determination of the SUSY breaking parameter $A_0$ and the stau mixing
 angle would require an improved knowledge of the polarized parton densities
 at small $x$.}
\end{figure}

The stau mixing angle depends directly on the universal soft SUSY breaking
mass scale $\Lambda$ in the GMSB model and on the trilinear coupling $A_0$
in the mSUGRA models. We test the sensitivity of the single-spin
asymmetry on these parameters by varying them within their allowed ranges
(see Figs.\ \ref{fig:2}-\ref{fig:4}). We note in passing that the reduced
value of $A_0=-300$ GeV proposed in Ref.\ \cite{spa} for the mSUGRA point
SPS 1a' leads in SUSPECT to a Higgs potential that is unbounded from below.

For the only existing polarized hadron collider RHIC, which will be
operating at a center-of-mass energy of $\sqrt{S}=500$ GeV in the near
future, and in the GMSB model with a light tau slepton, we show the
single-spin asymmetry in Fig.\ \ref{fig:2} as a function of the cosine of
the stau mixing angle. The
asymmetry is quite large and depends strongly on the stau mixing angle.
However, very large values of $\cos\theta_{\tilde{\tau}}$ and stau masses
below 52 GeV may already be excluded by LEP \cite{Barate:1998zp}, while
small values of $\cos\theta_{\tilde{\tau}}$ may be unaccessible at RHIC due
to its limited luminosity, which is not expected to exceed 1 fb$^{-1}$.
Polarization of the proton beam will also not be perfect, and the calculated
asymmetries should be multiplied by the degree of beam polarization $P_L
\simeq 0.7$. The uncertainty introduced by the polarized parton densities
increases considerably to the left of the plot, where the stau mass 41 GeV
$\leq m_{\tilde{\tau}}\leq$ 156 GeV and the associated values of the parton
momentum fractions $x_{a,b}\simeq2m_{\tilde{\tau}}/\sqrt{S}$ become large.

As mentioned above, the SM background cross section can be reduced by
imposing an invariant mass cut on the observed tau lepton pair, {\it e.g.}
of 2$\cdot$52 GeV. While the cross section of 0.13 pb is then still two
orders of magnitude larger than the SUSY signal cross section of 1 fb, the
SM asymmetry of -0.04 for standard polarized parton densities or -0.10 for
the valence-type polarized parton densities can clearly be distinguished
from the SUSY signal due to its different sign.

While the variation of the parameter $\Lambda$ in the GMSB model introduced
not only a variation of the stau mixing angle, but also of the stau mass,
variation of the parameter $A_0$ in mSUGRA leaves the stau mass almost
invariant. In the SPS 1a' model, its value varies only between 114 and 119
GeV, and the corresponding unpolarized cross section at the Tevatron is
nearly constant ($\sim 5.8\pm0.5$ fb).

%
%

This would make an asymmetry measurement at an upgraded Tevatron extremely
valuable, as one can see in Fig.\ \ref{fig:3}. The predicted asymmetry is
very sizeable in the entire viable SUSY parameter range, and it depends
strongly on the parameter $A_0$ and the stau mixing angle. Unfortunately,
the parton density uncertainty is still large, but it will be reduced
considerably in the future through more precise measurements at the COMPASS,
HERMES, PHENIX, and STAR experiments. As a recent experimental study
demonstrates, events with tau lepton pairs or tau leptons with associated
missing energy larger than 20 GeV can be identified with the CDF-II detector
in events with hadronic tau decays \cite{Anastassov:2003vc}.

The SM background cross section after an invariant mass cut of 2$\cdot$119
GeV is 0.16 pb and thus only about 25 times larger than the
SUSY signal. As in the case of RHIC, the SM asymmetry of -0.09 (for both
polarized parton densities) would be clearly distinguishable due to its
opposite sign.

For the LHC, where first feasibility studies of tau slepton identification
with the ATLAS detector \cite{Hinchliffe:2002se} and tau tagging with
the CMS detector \cite{Gennai:2002qq} have recently been performed
and SUSY masses should in general be observable up to the TeV range, we
choose an mSUGRA model with elevated scalar mass $m_0=400$ GeV and a large
value of $\tan\beta=50$, which enhances mixing in the stau sector. While
the predicted asymmetry for a possible polarization upgrade of the LHC in
Fig.\ \ref{fig:4} is slightly smaller than in the previous two cases, it
is still comfortably large and has again the opposite sign with respect to
the SM asymmetry of -0.02 (for both polarized parton densities). The
dependence of the asymmetry on the stau mixing angle is, however, also
reduced, while the uncertainties from the polarized parton densities, which
are not yet well known at the small $x$ values relevant for the large LHC
center-of-mass energy, are quite enhanced.

\section{Conclusion}
\label{sec:4}

In this Letter, we have presented a new calculation of cross sections and
asymmetries for slepton pair production through neutral and charged
electroweak currents in polarized hadron collisions. Our analytical results
are valid for general slepton masses and include the mixing of the left- and
right-handed interaction eigenstates relevant for third generation sleptons.
They confirm and extend in these respects an earlier pioneering calculation
\cite{Chiappetta:1985ku}.

Numerically, we have studied in detail the dependence of the longitudinal
single-spin asymmetry on the tau slepton mixing angle for pair production of
the lighter tau slepton mass eigenstate. Its physical mass and the mixing
angle at the electroweak scale have been calculated with the help of
renormalization group equations after imposing restricted sets of SUSY
breaking parameters at the unification scale.

The determination of these parameters in measurements of the longitudinal
single-spin asymmetry at the only existing polarized $pp$ collider RHIC
was found to be difficult due to its limited center-of-mass energy and
luminosity, even in a gauge mediated SUSY breaking model with a very light
tau slepton.

In contrast, a polarization upgrade for the proton beam of the Tevatron
would give direct access to the trilinear coupling $A_0$ in a typical
minimal supergravity model, independently of the tau slepton mass and the
unpolarized cross section.

At the LHC, where larger masses are easily accessible and where we have
studied an alternative minimal supergravity model with enhanced tau slepton
masses and mixings, the sensitivity of the longitudinal single-spin
asymmetry to the mixing angle and the trilinear coupling $A_0$ is found to
be reduced and hampered by a large uncertainty from the not well-known
polarized parton densities at small values of their longitudinal momentum
fractions in the proton.

For all colliders, an asymmetry measurement would allow for a
straightforward discrimination of the SUSY signal from the associated SM
background of tau lepton pair production due to the opposite sign of SUSY
and SM asymmetries.

\section*{Acknowledgments}
We thank A.\ de Roeck and G.\ Polesello for interesting discussions and
encouragement at the Frascati meeting of the Euro-GDR on {\em
Supersymmetry}.
This work was supported by a postdoctoral and a Ph.D.\ fellowship of the
French ministry for education and research.

\appendix

\section{Slepton Mixing}
\label{sec:a}

The (generally complex) soft SUSY-breaking terms $A_l$ of the trilinear
Higgs-slepton-slepton interaction and the (also generally complex)
off-diagonal Higgs mass parameter $\mu$ in the MSSM Lagrangian induce
mixings of the left- and right-handed slepton eigenstates $\tilde{l}_{L,R}$
of the electroweak interaction into mass eigenstates $\tilde{l}_{1,2}$. The
slepton mass matrix \cite{Haber:1984rc}
\beq
 {\mathcal M}^2 =
 \lr\begin{array}{cc}
  m_{LL}^2+m_l^2  &
  m_l m_{LR}^\ast \\
  m_l m_{LR}      &
  m_{RR}^2+m_l^2
 \end{array}\rr
\eeq
with
\bea
 m_{LL}^2&=&(T_l^3-e_l\sin^2\theta_W)m_Z^2\cos2\beta+m_{\tilde{L}}^2,\\
 m_{RR}^2&=&e_l\sin^2\theta_W m_Z^2\cos2\beta+\left\{\begin{array}{l}
 m_{\tilde{\nu}}^2\hspace*{4.8mm}{\rm for~sneutrinos},\\
 m_{\tilde{l}}^2\hspace*{4.5mm}{\rm for~charged~sleptons},\end{array}\right.\\
 m_{LR}  &=&A_l-\mu^\ast\left\{\begin{array}{l}
 \cot\beta\hspace*{5mm}{\rm for~sneutrinos}\\
 \tan\beta\hspace*{4.5mm}{\rm for~charged~sleptons}\end{array}\right.
\eea
is diagonalized by a unitary matrix $S$, $S {\mathcal M}^2 S^\dagger=
{\rm diag}\, (m_1^2,m_2^2)$, and has the squared mass eigenvalues
\beq
 m_{1,2}^2=m_l^2+{1\over 2}\lr m_{LL}^2+m_{RR}^2\mp\sqrt{(m_{LL}^2-
 m_{RR}^2)^2+4 m_l^2 |m_{LR}|^2}\rr.
\eeq
For real values of $m_{LR}$, the slepton mixing angle $\theta_{\tilde{l}}$,
$0\leq\theta_{\tilde{l}}\leq\pi/2$, in
\beq
 S = \lr \begin{array}{cc}~~\,\cos\theta_{\tilde{l}} &
                              \sin\theta_{\tilde{l}} \\
                             -\sin\theta_{\tilde{l}} &
                              \cos\theta_{\tilde{l}} \end{array} \rr
 \hspace*{3mm} {\rm with} \hspace*{3mm}
   \lr \begin{array}{c} \tilde{l}_1 \\ \tilde{l}_2 \end{array} \rr =
 S \lr \begin{array}{c} \tilde{l}_L \\ \tilde{l}_R \end{array} \rr
\eeq
can be obtained from
\beq
 \tan2\theta_{\tilde{l}}={2m_lm_{LR}\over m_{LL}^2-m_{RR}^2}.
\eeq
If $m_{LR}$ is complex, one may first choose a suitable phase rotation
$\tilde{l}_R'=e^{i\phi}\tilde{l}_R$ to make the mass matrix real and then
diagonalize it for $\tilde{l}_L$ and $\tilde{l}_R'$. $\tan\beta$ is the
(real) ratio of the vacuum expectation values of the two Higgs fields. The
soft SUSY-breaking mass terms for left- and right-handed sleptons are
$m_{\tilde{L}}$ and $m_{\tilde{\nu}}$, $m_{\tilde{l}}$, respectively.



\end{document}